# Characteristics of LIS Research Articles Affecting Their Citation Impact

Authors:

Kalervo Järvelin, Tampere University

Yu-Wei Chang, National Taiwan University

Pertti Vakkari, Tampere University

Address:

Järvelin & Vakkari:

Faculty of Information Technology and Communication Sciences | Communication Sciences, Tampere University, FI-33014 Tampere University, Finland

Chang:

Dept. LIS, NTU, 1, Sec.4, Roosevelt Road, Taipei 10617, Taiwan (R.O.C)

Tel:

+358 50 4034353          (Järvelin)
+886-2-3366-1743          (Chang)
+358 50 5288354          (Vakkari)

Email:

yuweichang2013@ntu.edu.tw
kalervo.jarvelin@tuni.fi
pertti.vakkari@tuni.fi

Corresponding author:

Vakkari is the author to receive correspondence and proofs.

# Abstract

The paper analyses the citation impact of Library and Information Science (LIS) research articles published in 31 leading international LIS journals in 2015. The main research question is: to what degree do authors' disciplinary composition in association with other content characteristics of LIS articles affect their citation impact? The impact is analysed in terms of the number of citations received and their authority, using outlier normalization and subfield normalization. The article characteristics analysed using quantitative content analysis include topic, methodology, type of contribution, and the disciplinary composition of their author teams. The citations received by the articles are traced from 2015 to May 2021. Citing document authority is measured by the citations they had received up to May 2021. The overall finding was that authors' disciplinary composition is significantly associated with citation scores. The differences in citation scores between disciplinary compositions appeared typically within information retrieval and scientific communication. In both topics LIS and computer science jointly received significantly higher citation scores than many disciplines like LIS alone or humanities in information retrieval; or natural sciences, medicine, or social sciences alone in scientific communication. The paper is original in allowing joint analysis of content, authorship composition, and impact.

## 1. INTRODUCTION

Due to the central status of citation impact analysis in research assessment, it is important to understand what affects citation impact of scholarly publications. Reasons to cite are well documented in the literature of bibliometrics (e.g., Weinstock, 1971; Tahamtan & Bornmann, 2019). These include paying homage to prior work, assessing current work, justifying claims, etc. However, these are rather *normative* than *descriptive reasons* for citing, and often not followed in research practice. Descriptive reasons include self-citation, cronyism, and language bias (Latour, 1987; Phelan, 1999). Their effect on citation impact is debatable. In either case the effect cannot always be analyzed based on the documents (or their bibliographic data) alone but require contextual information.

In the present work, we shall analyze citation impact by factors for which data can be collected from the documents. We divide these factors into *contextual factors* and *content factors*. Examples of the former are author names, journal titles, institution names, country names, years of publication etc. These are easy to collect for automatic informetric analysis and have been thoroughly studied in the literature (e.g., Phelan, 1999). These factors do not inform about the content of study, directly. The content factors in scholarly publications report about the study topic, research problem, the approach to the problem, the data collection method, the type of analysis, and the type of contribution. These factors are present in document texts but not trivial to collect for impact analysis and have not been thoroughly studied for their impact on citations (Yan, 2015). Due to the central role of the latter in research design, execution, reporting, and assessment, we expect that they are associated with the impact of research. A negative answer would indicate a liberal appreciation of all types of research within the research field. At least, the explanation would lie beyond the study topic and type of methodology. On the other hand, a positive answer would indicate that some combinations of research design and type of methodology have greater impact than others among researchers.

At the level of publications, citation impact means counting the citations a given publication has received and expressing the result as some index value. One must be careful with interpreting citation counts because they are typically skewed and require adjustment for comparable results: the sizes of the fields must be considered (e.g., Phelan, 1999; Waltman, 2016). Various normalized indices of citation impact have been proposed. They adjust field sizes, time factors (publication and citation windows; aging of citations), publication types, or citation weights (e.g., Järvelin & Persson, 2008; Waltman, 2016). Some indices measure the impact of individual publications, e.g., the Field-Weighed Citation Impact, while others measure the aggregate impact of a set of publications from the same source, like the H-index and its derivatives (e.g., Waltman, 2016). In the present paper we analyze the citation impact of articles in a narrow publication window (2015) of a single field (LIS) in a restricted citation window (2015 – May 2021). Therefore, the conditions are comparable for all articles and normalization for publication year, publication type, or citation year are not relevant.

The *topic* of a publication in LIS may affect its impact because LIS consists of several quite different research traditions which are of interest to different academic and professional audiences. Vakkari (1994) discusses the nature of LIS at a theoretical level, and Jarvelin and Vakkari (1990, 1993, 2021) and Tuomaala et al. (2014) on the level of empirical characteristics. These studies inform about the knowledge production side in a discipline, but only indirectly about the quality of research. While there are aggregate indicators of publication forum quality

(such as *JIF*), these cover the entire forums as blocks and do not report on impact at the level of topics covered by the forums. The overall *research strategy* and *type of contribution* may also affect the citation impact of a publication because some of the formers are clearly more popular than others (e.g., Jarvelin & Vakkari, 1990; 1993; 2022; Tuomaala et al., 2014).

The *disciplinary composition* of the article author teams likely affects the choice of research topics, definition of research problems, and methodological choices (Vakkari, Chang & Järvelin 2022a,b). This means that the produced contributions draw on varied literatures and may also contribute to them, and therefore interact with the citation impact.

The main concepts of the study are explained in Figure 1. The unit of analysis is a cited LIS journal article (primary article in the figure). Its impact on knowledge production, citation impact, is expressed through the number of secondary documents citing it. The citing documents, however, have varying *authority* based on the citation count by tertiary documents. Instead of using plain citation count to measure the impact of a LIS article, we calculate its *impact score* on the basis of its authority-weighed citations. Our aim is to find out whether the cited LIS article content characteristics: topic, research strategy, contribution type, or author team disciplinary composition are associated with the variance of its impact score.

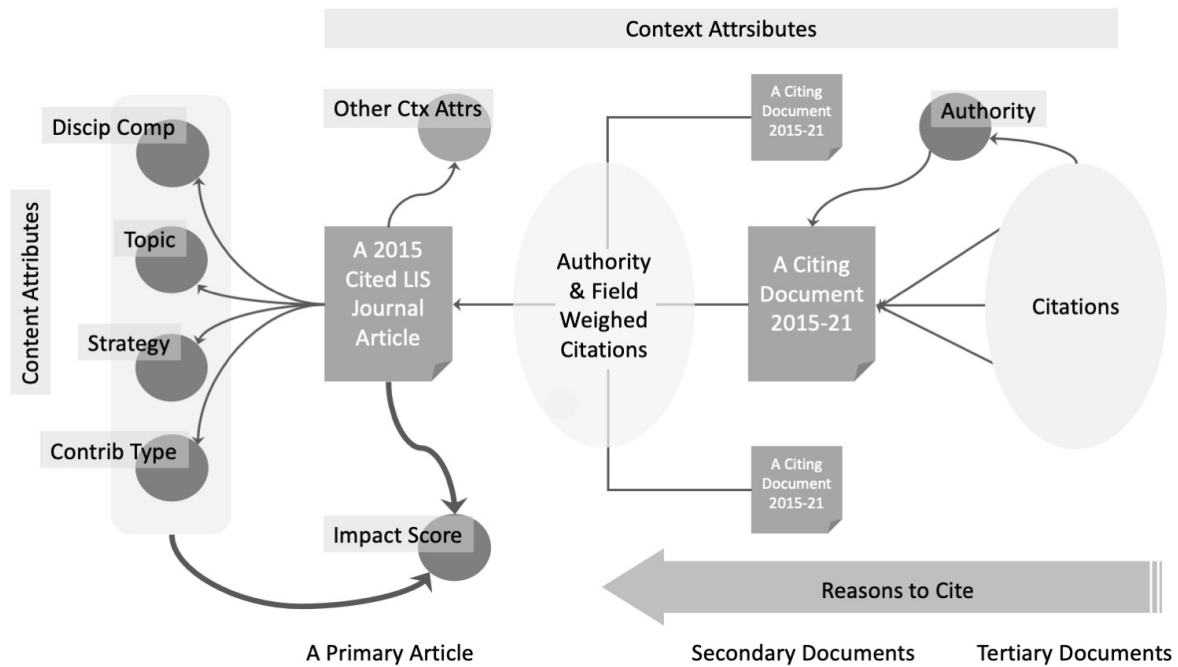

**Figure 1.** The main concepts of the study

The content attributes, on the left, are derived through content analysis of the primary articles. The context attributes, in the middle and on the right in Fig. 1, focus on citations, their authority and normalization through field weighting, producing the impact score of the primary article.

The findings of the present study contribute to a) the understanding the nature of LIS, b) designing research programs with 'right' combination of knowledge, c) planning doctoral programs and recruiting doctoral students, and d) individual scholars in planning their careers. Further, contributions are offered to the e) empirical and g) methodological body of knowledge of informetrics.

## 2 PRIOR WORK
### 2.1 Methods of Empirical Study of LIS

Identifying main topics and their changes has long been a focus in LIS research. Studies have explored LIS-related topics applying several methods, including content analysis (Järvelin & Vakkari, 1990, 1993, 2022; Tuomaala et al., 2014), co-citation analysis (Åström, 2007; Hou et al., 2018), bibliographical coupling (Chang et al., 2015), keyword analysis (Liu & Yang, 2019), subject index term analysis (Blessinger & Frasier, 2007), co-word analysis (Mokhtarpour & Khasseh, 2021), latent Dirichlet allocation modeling (Figuerola et al., 2017; Han, 2020), also in various combinations (Chang et al., 2015). Content analysis is a traditional method for deep analysis. Through expert judgment, it can reveal multiple characteristics of research articles. For example, the classification framework by Jarvelin and Vakkari (1990) consists of topics, methods, viewpoints, and strategies, and is frequently used by other researchers (Hider & Pymm, 2008; Lund & Wang, 2021; Ma & Lund, 2021).

### 2.2 Journal Articles as Data Source

Journal articles have been typical units of observation in scientometric analyses. The articles observed have inherited their disciplinary attributes from the journal, in which they have been published. This has at least three drawbacks. All articles in a journal neither represent the discipline of the journal as classified in, e.g., the Web-of-Science, nor are they all research articles. For instance, Järvelin and Vakkari (2022) have shown that in 2015 in 31 major LIS research journals 13% of articles did not belong to LIS, and 7% were not research. In 2005 the share of non-research articles was larger, 30% (Tuomaala et al., 2014). The third limitation is that journal-based article characterization conceals the sub-field (topical) variation of articles, which influence citations and impact (Yan, 2015). Thus, journal-based analysis may reduce the validity of findings by blurring the boundaries between disciplines, including non-research articles in the analysis, and ignoring the variation in impact between disciplines' sub-fields. In the present paper, we avoid these pitfalls.

## 2.3 LIS and non-LIS authors in LIS publications

The presence of non-LIS authors in LIS research is increasing, causing non-LIS authors dominate the LIS research landscape (Chang, 2019; Chang & Huang, 2012; Urbano & Ardanuy, 2020; Vakkari et al. 2022a,b). Author affiliation information is usually used to determine authors' disciplines, which are then used to identify articles produced through interdisciplinary collaboration. Studies have verified the contributions of non-LIS authors from numerous disciplines to LIS publications and the substantial proportion of non-LIS authors who were affiliated with computer science–related departments and institutes (Chang, 2018; Vakkari et al. 2022a). Chang (2019), examining articles of 75 LIS journals in 2015, also reported that approximately 70% of LIS journals were dominated by non-LIS authors. Chang (2018) verified that non-LIS and LIS authors have different topic preferences.

Vakkari et al. (2022b), using the dataset of the present paper, found that most articles in LIS are contributed by non-LIS authors (57%). Still, LIS scholars have a clear majority in research on L&I services and institutions (68%), while external scholars dominate information retrieval (73%) and scientific communication (scientometrics, 69%). Vakkari et al. (2022a), using the same dataset, found that the share of LIS was one third, while computer science contributed one fifth and business / economics one sixth. The latter dominate the contributions in information retrieval, and scientific communication indicating strong influences in LIS.

## 2.4 Citation impact in LIS

Non-LIS authors have considerable influence on the evolution of LIS because of their preferences pertaining to topics and methods (Vakkari et al. 2022a,b), and thus, we assessed research influence by topic and author discipline and examined citation counts. Citation counts of scientific publications are widely applied in scientometrics as quick indicators of influential publications. However, some researchers have argued that the use of citation-related indicators is inappropriate for assessing research influence (MacRoberts & MacRoberts, 2018).

Nevertheless, several studies (Jirschitzka et al., 2017; So, 1998) indicate positive correlation between citation count and expert judgment.

Regardless of the factors that proliferate citations, highly cited articles receive attention from multiple fields. However, no study has compared the differences in the scientific influence of LIS-related topics. In LIS research, highly cited articles and their characteristics (e.g., topics) have been explored. Blessinger and Hrycaj (2010) identified 32 highly cited articles that were most frequently cited by LIS articles that were published in 10 LIS journals between 1994 and 2004. After classifying subject index terms in the articles into seven categories, they discovered that the most highly cited articles focused on librarianship and users (68%), followed by technology (22%). Sahoo et al. (2020) analyzed the topics of 166 highly cited articles in four LIS journals. These were determined through content analysis. Among the 14 topics identified, research impact measurement and research collaboration were the most frequent topics (26%) and had the most influence (29% of the total citation count). Given that past studies have mainly focused on specific characteristics embedded in LIS publications, the present study aimed to determine the connections among multiple characteristics, such as research topic, research strategy, type of contribution, and author discipline.

## 2.5 Citation impact metrics

Publication-based impact metrics are relevant for the present study. Care is needed when interpreting citation counts because they are typically skew, and the datasets may cover larger or smaller disciplines. The former causes outliers to distort the findings (Phelan, 1999), and the latter the number of citations that a publication can be expected to receive (e.g., Waltman, 2016). Various field-weighted, or normalized, versions of citation impact have been proposed to mitigate this (e.g., Waltman, 2016). Citation counts may also be normalized for publication type and year. Moreover, one may argue that citations by a high-impact publications weigh more than ones by a low-impact publication (e.g., Waltman, 2016). The credit accrued by citations may also be discounted by time of citing. For this, the Discounted Cumulative Impact index allows devaluing old citations (Järvelin & Persson 2008; Ahlgren & Järvelin 2010). Our data represents a single field, LIS, and scholarly journal articles published in its top journals in a single year, 2015. Therefore, we apply logarithmic smoothing of outliers, and subfield-based field normalization in impact analysis.

## 3 METHODS

## 3.1 Research questions

The main research question is: Which content factors of research articles in LIS are associated with impact scores received? We elaborate the analysis through the following sub-questions:

- RQ1: What are the effects of disciplinary composition of authors on impact scores?
- RQ2: What are the combined effects of disciplinary composition and each of the other content factors: (a) research topic, (b) research strategy, and (c) type of contribution on impact scores?

Figure 2 illustrates the study design. The cloud on the left represents research work, detailing four of its theoretical features. Researchers work in the context of some discipline, as part of a research program, with project level aims following a strategy within which some methods are applied. When reporting the project as an article, these features appear as text content. We operationalize them, partially, as the content factors disciplinary composition, topic, strategy, and contribution type. We then analyze their main and combined effects on article impact scores.

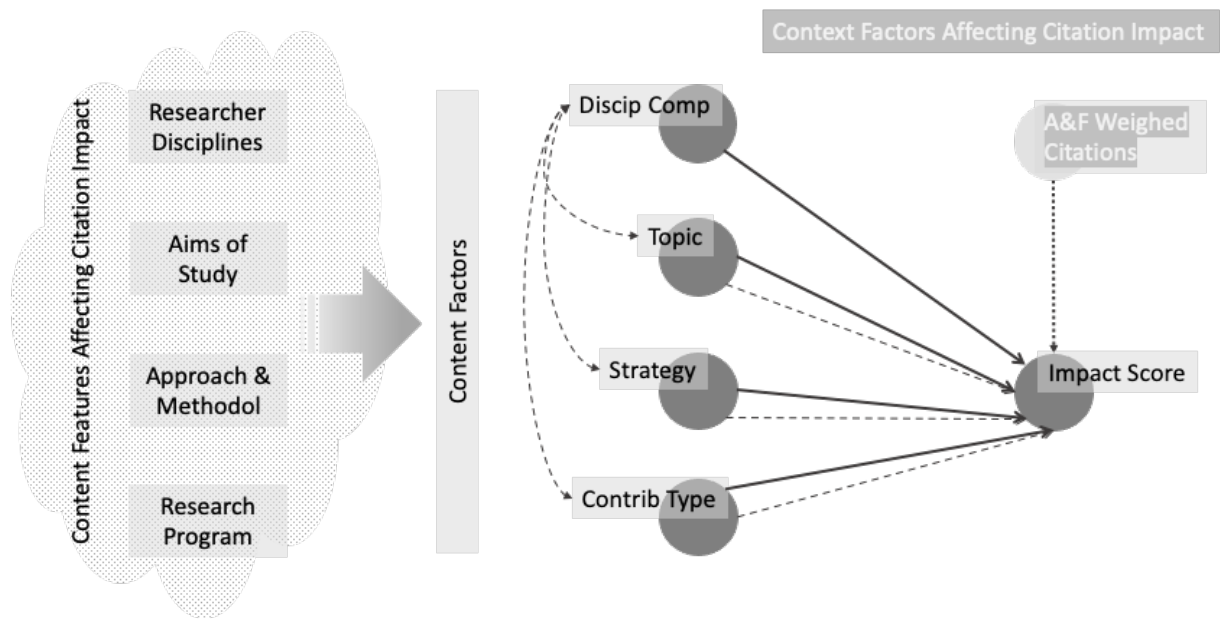

**Figure 2.** The study design. Solid arrows represent main effects, dashed arrows combined ones, and dotted arrows constants.

## 3.2 Data collection

The research process is illustrated in Figure 3 and the dimensions of the data set are summarized in Table 1. The data set consists of:

- a quantitative intellectual content analysis of articles published in scholarly LIS journals in 2015 (Järvelin & Vakkari, 2022) collected in the First phase,
- data on the disciplinary composition of the author teams (Vakkari & Chang & Järvelin, 2022a,b) collected in the Second phase, and
- data on the ~25K citations to the LIS article set between 2015 and May 2021 derived from Scopus in the Third phase reported in the present paper.

The content analysis of the LIS article set focuses on research topics, methods, and types of contributions. Article authors were assigned to one of eight disciplinary categories as indicated by their affiliations, forming the disciplinary composition of each author team. Finally, the citations received by the articles were traced from 2015 to May 2021. Citing document authority was measured by the citations they had received up to May 2021. Next, we describe the data in more detail.

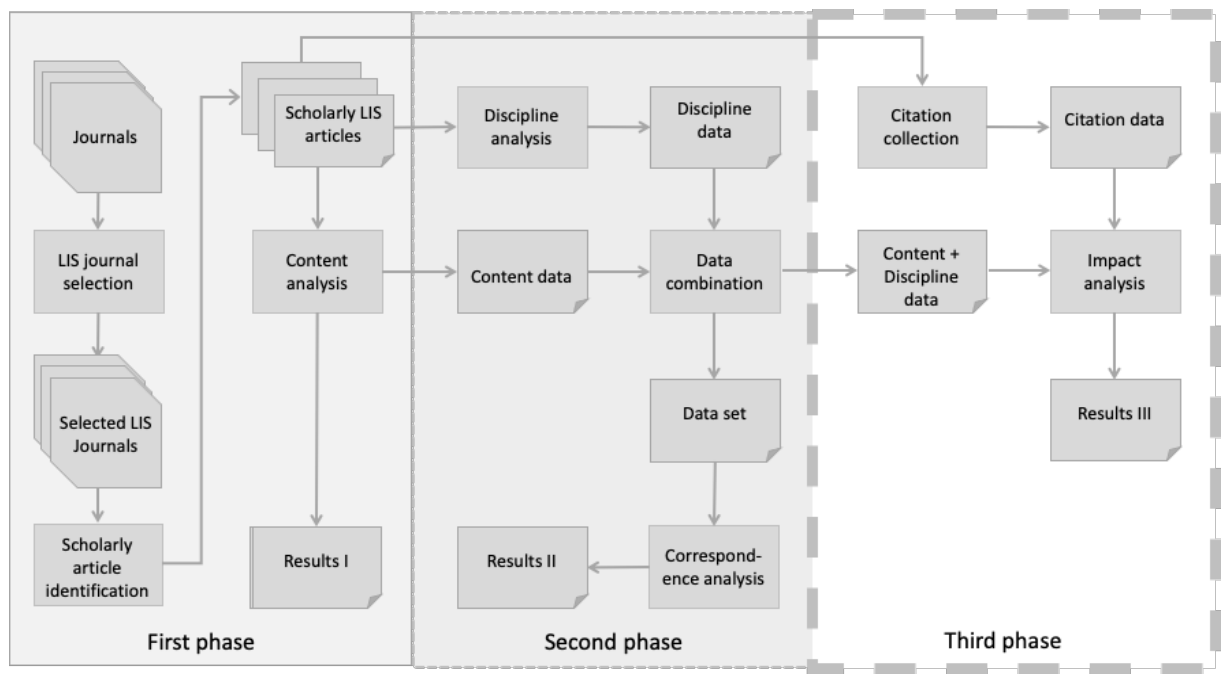

**Figure 3.** The data collection, preparation, and analysis process (based on Vakkari & Chang & Järvelin, 2022a)

| Object / Attribute | Value |

| Journals | |
|---|---|
| Volume | 2015 |
| Unit of observation | A journal |
| Total number of titles | 31 |
| **Articles** | |
| Unit of observation | An article |
| Total number | 1514 |
| - excluding non-LIS | 1322 |
| - excluding non-research | 1210 |
| Content dimensions | 3 |
| Classifiers, equal shares | 2 |
| **Disciplinary Compositions** | |
| Unit of observation | The pair (article, discipline) |
| Total number | 1533 |
| Content dimensions | 3 |
| Classifiers | 1 |
| **Citations** | |
| Unit of observation | The pair (cited, citing doc) |
| Total number | 24965 |
| Content dimensions | 1 |
| Classifiers | - |

**Table 1.** Dimensions of data

## 3.3 Content analysis

Table 2 introduces the content analytical variables used in the present study. Appendix II defines them by listing their classes. The classification of research *topics* was used at the level of four main research topics for analysis: *L&I context and services*, *information retrieval (IR)*, *information seeking*, and *scientific communication*. The research topic class *Non-LIS research* was coded as well but excluded from the analyses.

| **Content analytic variables** (see Järvelin & Vakkari, 2022) | |
|---|---|
| **Name** | **Explanation** |
| LIS topic | The focus of an article, e.g., information seeking, expressed as a main topic |
| Research strategy | Indicates the overall combination of data-collection and analysis methods of the study |
| Type of contribution | Indicates empirical, theoretical, methodological, constructive, etc. contribution type of a study |
| **Discipline analytic variables** | |
| **Name** | **Explanation** |
| Discipline | Gives each unique discipline name based on an article's co-authors' affiliations. |
| No. of disciplines | Indicates the number of unique disciplines contributing to an article |

**Table 2.** Content and discipline analytic variables of the data set

For increasing the degrees of freedom in the analysis we merged classes of some variables. In research strategies historical, and evaluation strategies were merged with other empirical strategies as other empirical strategies; citation analysis was merged with other bibliometric strategy as citation analysis; verbal argumentation and concept analysis were merged as conceptual strategy; literature review and bibliographic strategy were merged with other strategy as other strategy (Appendix II).

Each article was classified under one content class for each content variable. Classification reliability was measured by Fleiss' *Kappa* (Table 3). The classification of *Main topic,* and *Topic* had good agreement, while *Research strategy* and *Type of contribution* had moderate.

| **Content analytic variables** | (**N = 32**) | | | |
|---|---|---|---|---|
| **Name** | **Kappa** | ***p*-value** | **No of Raters** | **Level** |
| LIS main topic | 0.73 | 0.000 | 2 | good |
| LIS topic | 0.62 | 0.000 | 2 | good |
| Research strategy | 0.53 | 0.000 | 2 | moderate |
| Type of contribution | 0.60 | 0.000 | 2 | moderate |
| **Discipline analytic variables (N = 40)** | | | | |
| **Name** | **Kappa** | ***p*-value** | **No of Raters** | **Level** |
| Discipline | 0.71 | 0.000 | 3 | good |
| No. of disciplines | 0.64 | 0.000 | 3 | good |

**Table 3.** Classification reliability (Fleiss' *Kappa*)

## 3.4 Encoding of authors' disciplines

Chang's (2018) method was used to identify each author's disciplinary affiliation. Its application to the present data set is described in Vakkari et al. (2022a). The main points are:

(1) Affiliations with LIS-related institutions were classified as LIS affiliations. (2) Other authors were classified with disciplinary affiliation in: Business-and-economics, computer science, engineering, humanities, medicine, natural sciences, and social sciences (see Appendix III). (3) The present study employed various reference sources and the Internet to identify some authors' affiliation when the information provided in the article was incomplete.

Each article was thereafter classified as having each distinct disciplinary affiliation. The number of authors representing a single discipline, or their order, was ignored. Thus the 1513 articles gave, after pruning off non-scholarly articles and non-LIS articles, a set of 1210 source articles. Some of the primary source articles lacked citation data producing a set of 1181 articles. The analysis is restricted to articles with less than three contributing disciplines (n=1145) to avoid the loss of degrees of freedom when combining disciplines. These articles cover 97 % of all articles. In addition, three articles were removed due to being outliers in citation counts.

## 3.5 Collecting Citation Data for Citing Document Authority

To assess the impact of a LIS article, we determined its impact score. This was based on the number of weighed citations that the article had received. Each citation is an elementary indication of impact and the greater the impact, the more there are citations and vice versa (see Figure 1). Moreover, the more authoritative the citing secondary documents are, the greater the impact of the primary LIS article is. We calculate a secondary document's citing authority by the count of citations the secondary document itself has received during 2015 – May 2021 according to the Scopus database. The number of citations approximates impact and does not consider the quality of the tertiary citing publications, the publication years of the secondary or tertiary documents, or various reasons of citing. The conditions are however the same for all citing documents, most importantly the primary article publication window (year 2015) and the secondary and tertiary document citation windows (2015 – May 2021). Therefore, the contribution of the factors in Figure 2 to the citation impact of a LIS article can be reliably analysed.

However, determining document authority by its raw citation count is problematic, see Figure 4. The range of citations received by the secondary documents was [0, 1397] and the distribution very skewed. Nearly 7,000 of the almost 25,000 secondary documents had no citations, and two thirds had six or less. We normalized the citation count in three ways: first smoothed them by taking logs; second by analysing them by the primary article main topic for control of field-size effects; and third by calculating the topic-specific relative to the average score as the citation impact of the primary article.

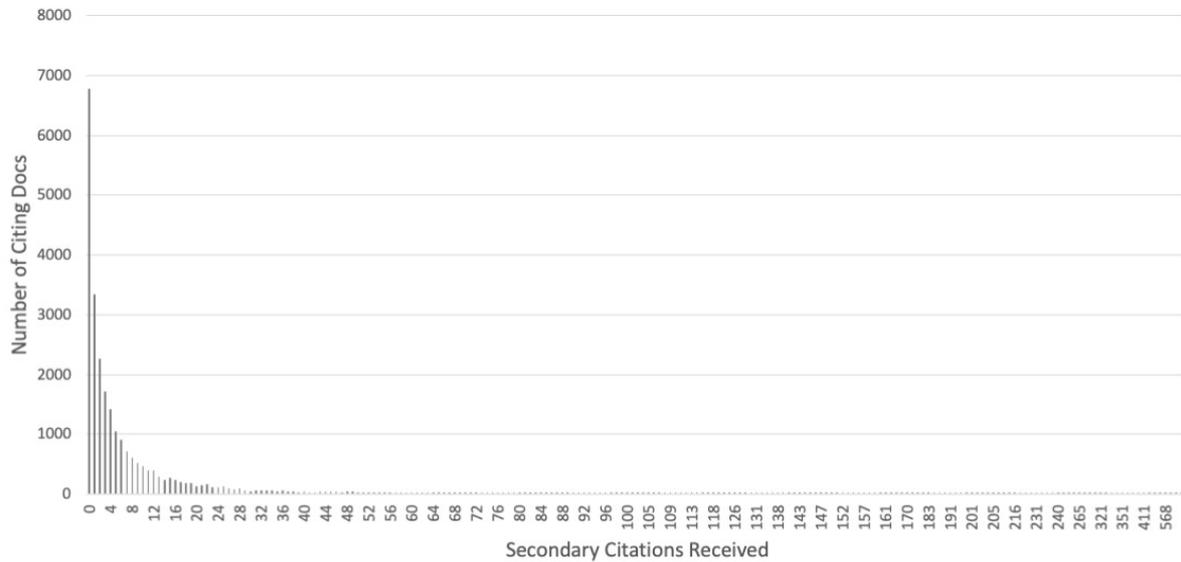

**Figure 4.** Determining authority: the distribution of citing secondary documents by the number of citations received. The x-axis is cut at 568 citations.

Smoothing by taking logarithms was intended to reduce the effects of outliers. We experimented with several log bases $b$ (1.5, 2, and 10) and found $b = 2$ a good compromise between preserving differences and effectiveness of smoothing of outliers. Let $cc_{sd}$ denote the citation count of a secondary document $sd$. The smoothed citation count $scc$ of $sd$ is calculated as:

$$scc_{sd} = \log_2(cc_{sd} + 1)$$

The "+1" is needed because $\log_b 0$ is undefined. The smoothed authority range becomes [0, 10.45]. Given a primary document $pd$ and a set of secondary documents $SD_{pd} = \{sd_{1,pd}, sd_{2,pd}, \ldots, sd_{n,pd}\}$ citing it, the authority-weighed citation impact of $pd$ is:

$$AWCI_{pd} = \sum\nolimits_{sd \in SD} scc_{sd} \; .$$

The $AWCI_{pd}$ counts all secondary documents $sd$ citing the article $pd$ and contributes to $pd$ the sum of smoothed authorities which can be zero if $sd$ has no citations ($cc_{sd} = 0$). However, $AWCI_{pd}$ does not provide field normalization. Given a set of primary articles $PD = \{pd_1, \ldots, pd_n\}$ representing a field, we use the average smoothed authority:

$$avg\text{-}AWCI_{PD} = \sum_{pd \,\in\, PD} AWCI_{pd} \big/ |PD|$$

to assign $pd$ the *Authority & Field-Normalized Impact Score, $AFNIS_{pd}$,* by the following function:

$$impactscore(pd, PD) = AWCI_{pd} \big/ avg\text{-}AWCI_{PD}$$

The range of the normalized impact scores is [0, z] where z is a real number. The average value of the scores is 1, while often z >> 1. Table 4 gives statistics of various citation counts for LIS and its subfields.

| Name | LIS subfield | | | | Total |
|---|---|---|---|---|---|
| | LIS Ctx | IR | ISeek | Scim | |
| Avg of Citations $cc_{pd}$ | 9.78 | 13.34 | 16.77 | 20.37 | 15.63 |
| Max of Citations $cc_{pd}$ | 320.00 | 106.00 | 184.00 | 463.00 | 463.00 |
| $avg\text{-}AWCI_{PD}$ | 15.98 | 26.89 | 30.23 | 43.34 | 30.95 |
| Max $AWCI_{PD}$ | 728.44 | 230.40 | 296.08 | 873.00 | 873.00 |
| Max $AFNIS_{pd}$ | 45.57 | 8.57 | 9.79 | 20.14 | 28.20 |
| N | 291 | 273 | 166 | 451 | 1181 |

**Table 4.** The effect of normalization on impact scores by LIS subfield.

## 3.6 Data analysis

The final data matrix for analysis was constructed by combining the encoding of the authors' disciplines with the citation and content analysis data (Figure 3).

## 4 FINDINGS

## 4.1 Disciplinary composition and impact scores

Discipline compositions had a significant association with impact scores (df=10, F=3.1, p<.001) (Table 5). The highest impact scores were received by authors in LIS collaborating with computer scientists (1.44), and in computer science collaborating with social scientists (1.46). The lowest impact scores produced articles authored by LIS scholars jointly with other disciplines (0.31), by scholars in natural sciences and medicine (0.56) and in humanities (0.54). Post hoc analysis (Tukey HSD) indicated that articles by authors in natural sciences and medicine received significantly lower impact scores compared to articles produced in collaboration with LIS and computer science, and in collaboration with computer science and social sciences.

| Disciplinary composition | Mean | Stddev | N |
|---|---|---|---|
| LIS | 0.88 | 1.24 | 397 |
| LIS & social sciences | 0.94 | 0.98 | 54 |
| LIS & computer science | 1.44 | 1.65 | 47 |
| LIS & others | 0.31 | 0.34 | 15 |
| Comp science | 1.07 | 1.36 | 232 |
| Comp science & social science | 1.46 | 1.90 | 43 |
| Comp science & others | 0.98 | 1.56 | 24 |
| Social sciences | 0.90 | 1.07 | 204 |
| Social sciences & others | 0.98 | 1.35 | 25 |
| Natural  sciences & medicine | 0.56 | 1.20 | 70 |
| Humanities | 0.54 | 0.87 | 31 |
| **Total** | **0.94** | **1.28** | **1142** |

[Legend: LIS = Library and Information Science; Social sciences = Social sciences, Business and Economics; Comp science = Computer science, Engineering; Others = Other disciplines]
**Table 5.** Impact scores by discipline compositions (RQ1)

The collaboration of LIS with computer science produced higher impact scores compared to other collaboration combinations of LIS. Also, the joint research efforts by computer science and social sciences lead to research results with high impact compared to many other disciplinary combinations. Thus, computer science is an influential actor in the field of LIS. However, it may be that the impact of disciplines varies between the sub-fields of LIS. Next, we analyze to what extent impact scores vary between disciplinary compositions and between the major topics of LIS.

## 4.2 Discipline, topic and impact scores

A two-way ANOVA of discipline compositions and topics was significant (df=42, F=1.9, p<.001). The main effect of neither topics (df=3, F=1.6, p=.18) nor discipline compositions (df=10, F=1.1, p=.38) on impact scores was significant, while the interaction effect (df=29, F=1.7, p=.012) was significant. Thus, disciplinary compositions influence differently on impact scores in various major topics. Next, we analyze this in more detail.

There were significant associations between disciplinary compositions and impact scores in some major topics. In IR there were no significant differences in impact scores between disciplinary groups (df=10, F=1.6, p=.10). However, detailed independent samples t-tests revealed that articles on IR authored solely by LIS scholars received significantly lower impact scores compared to articles authored by representants of computer science (Comp Sci) (df= 142.1, t=2.9, p=.004), and articles created jointly by scholars of LIS and computer science (df=63, t=2.4, p=.021). Humanities (Hum) received significantly lower impact scores compared to joint articles by LIS and computer science (df=21, t=3.0, p=.008) and computer science with social sciences (Soc Sci) df=25.4, t=2.5, p=.021).

Within information seeking disciplinary combinations differentiated significantly impact scores (df=9, F=4.8, p=.017). A post hoc analysis (Tukey HSD) indicated that articles produced by LIS scholars solely received significantly lower impact scores compared to joint articles by computer science and social sciences.

Within scientific communication disciplinary combinations significantly differentiated impact scores (df=10, F=2.8, p=.002). Independent samples t-tests revealed that articles authored by scholars in natural sciences and medicine (Nat Sci & Med) received significantly lower impact scores compared to LIS (df=99.3, t=3.5, p<.001), LIS in collaboration with social sciences (df=21.9, t=2.6, p=.018), LIS in collaboration with computer sciences (df=21.7, t=2.8, p=.01), computer sciences (df=73.4, t=3.1, p=.003) and social sciences (df=178.1, t=4.3, p<.001). Joint articles by scholars in LIS and computer science scored significantly higher compared to articles by social scientists (df=25.9, t=2.2, p=.035) and articles authored by scholars in computer science and other disciplines (Others) (df=23.5, t=2.3, p=.028).

Significant disciplinary differences within topics are summarized in Table 6.

| Topic | Significant differences |
|---|---|
| Information retrieval | • LIS < Comp Sci, LIS & Comp Sci<br>• Hum < LIS & Comp Sci, Comp Sci & Soc Sci |
| Information seeking | • LIS < Comp Sci & Soc Sci |
| Scientific communication | • LIS & Comp Sci > Soc Sci, Comp Sci & Others<br>• Nat Sci & Med < LIS, LIS & Soc Sci, LIS & Comp Sci, Comp Sci, Soc Sci |

**Table 6.** Significant differences in impact scores between disciplinary compositions within topics (RQ2a)

## 4.3 Discipline, research strategy and impact scores

A two-way ANOVA of disciplinary compositions and research strategies was significant (df=91, F=1.6, p<.001). The main effect of disciplinary composition (df=10, F=1.6, p=.10) and of research strategy (df=9, F=1.8, p=.07) were not significant, while their interaction effect was significant (df=72, F=1.4, p=.012).

There were significant associations between disciplinary compositions and impact scores within some research strategies. Within survey strategy, disciplinary combination differentiated impact scores significantly (df=10, F=2.5, p=.008). A post hoc analysis (Tukey HSD) indicated that joint articles by computer science and social sciences produced significantly higher impact than LIS, LIS and social sciences, computer science, social sciences, and natural sciences and medicine.

Citation analysis strategy (df=10, F=3.0, p=.09) did not significantly differentiate the impact scores of disciplinary groups. However, a post hoc analysis (Tukey HSD) indicated that within citation analytic strategy articles authored in collaboration by LIS and computer science produced significantly higher impact scores compared to natural sciences and medicine.

Within conceptual research strategy disciplinary compositions significantly differentiated impact scores (df=6, F=6.2, p<.001). A post hoc analysis (Tukey HSD) revealed that natural sciences and medicine produced significantly higher impact scores compared to LIS, LIS and social sciences, computer science, social sciences and humanities. In addition, articles authored in collaboration by computer science and social sciences produced significantly higher impact scores compared to articles authored by sole LIS scholars alone.

Significant disciplinary differences within research strategies are summarized in Table 7.

| Research strategy | Significant differences |
|---|---|
| Survey | • Comp Sci & Soc Sci > LIS, LIS & Soc, Comp Sci, Soc Sci, Nat Sci & Med |
| Citation analysis | • LIS & Comp Sci > Nat Sci & Med |
| Conceptual | • Nat Sci & Med > LIS, LIS & Soc Sci, Comp Sci, Soc Sci> Hum<br>• LIS < Comp Sci & Soc Sci |

**Table 7.** Significant differences in impact scores between disciplinary compositions within research strategies (RQ2b)

We have shown that there are some differences in impact scores between disciplinary groups both within topics and within research strategies. It may be that those disciplinary groups are more skillful in applying these strategies for solving research problems within certain topics thus producing higher impact scores. Our data is not large enough to directly elaborate this hypothesis. However, if there are no differences in impact scores in topics by research strategies, this suggests that there are no superior strategies as such to solve research problems in those topics. Combined with our previous results this would suggest indirectly that in the hands of some disciplines certain strategies would create results that produce higher impact scores. To test this hypothesis, we run a two-way ANOVA of topics and research strategies on impact scores, which was not significant (df=39, F=1.1, p=.28). This corroborates our conjecture that some disciplinary groups apply certain research strategies in some topics in a manner that produces high impact scores.

### 4.4 Discipline, the type of contribution and impact scores

A two-way ANOVA of disciplinary compositions and the type of contribution was significant (df=69, F=2.6, p<.001). The main effects of disciplinary composition (df=10, F=2.5, p=.006) and the type of contribution (df=7, F=4.9, p<.001) were significant as well as their interaction effect (df=52, F=2.3, p<.001). A post hoc analysis (Tukey HSD) showed that explanatory contributions received significantly higher impact scores compared to descriptive, theoretical, and methodological contributions.

There were significant associations between disciplinary composition and impact scores within some contribution types. Disciplinary combinations differentiated impact scores significantly (df=10, F=2.6, p=.002) within descriptive research type. A post hoc analysis (Tukey HSD) indicated that joint articles by LIS and computer science received significantly higher impact scores compared to LIS, natural sciences and medicine, and humanities.

Disciplinary composition differentiated impact scores significantly within methodological contributions (df=9, F=2.1, p=.042). LIS received significantly higher impact scores in the production of methodological contributions compared to natural sciences and medicine (df=17.2, t=2.2, p=.044).

Significant differences in impact scores within contribution types are summarized in Table 8.

| Contribution type | Significant differences |
|---|---|
| Descriptive | • LIS & Comp Sci > LIS, Nat Sci & Med, Hum |
| Methodological | • LIS > Nat Sci & Med |

**Table 8.** Significant differences in impact scores between disciplinary compositions within contribution types (RQ2c)

## 5 DISCUSSION

### 5.1 Main Findings

We have analyzed the citation impact of research articles through their *content factors* in the year 2015 batch of articles of leading LIS journals. We focused on the article authors' discipline, the topic, the methodology, and type of contribution claimed. These factors have a central role in research design, execution, reporting, and assessment. We investigated to what extent authors' disciplinary composition in connection to major topic, research strategy, and type of contribution – are associated with the impact of research. Impact was operationalized as the impact score for each article, and was based on the number of citations, and their authority, that the article received.

The collaboration of computer science either with LIS or social sciences led to significantly higher impact scores compared to natural sciences and medicine. These combinations produced highest impact scores, while humanities, natural sciences and medicine, and LIS in collaboration with other disciplines produced the lowest impact scores. These findings reveal the essential role of computer science in LIS for creating influential research results. They also indicate that it is profitable for scholars in LIS to collaborate with computer scientists for creating impactful research.

The differences in impact scores between disciplinary compositions appeared typically within information retrieval and scientific communication. In both topics LIS and computer science jointly received significantly higher impact scores than many other compositions like LIS alone or humanities in information retrieval; or natural sciences and medicine, or social sciences alone in scientific communication. Computer science in collaboration with social sciences produced high impact scores in information retrieval, and significantly higher impact scores in information seeking than LIS alone.

The composition of computer science and social sciences produced significantly higher impact scores in the use of survey research strategy compared to many disciplines like LIS, computer science or social sciences alone. This is likely associated with the high impact scores produced by this disciplinary combination in information seeking. Järvelin and Vakkari (2022) have shown that survey is the most popular research strategy in information seeking. Thus, the joint effort by computer science and social sciences in solving problems of information seeking by skillfully applying survey leads to productive results.

Reflecting the higher impact in scientific communication, LIS with computer science produced significantly higher impact scores in the use of citation analytic research strategy.

Within the descriptive contribution type, LIS with computer science received the highest impact, and within the methodological type, LIS alone was the most effective.

## 5.2 Interpreting the associations

There were some patterns in disciplinary differences between topics, research strategies and contribution types. Within information retrieval the combination of LIS and computer science, and computer science alone received significantly higher impact scores, whereas humanities significantly lower scores, compared to some other compositions. Mathematical, experimental and system analytic strategies were in 2015 the three most common strategies within information retrieval (Järvelin & Vakkari 2022). Our results show that within mathematical strategy, computer science and LIS in collaboration with computer science produced somewhat higher impact scores than other disciplinary compositions. Within system analytic strategy, computer science and LIS with computer science received notably higher impact scores compared to LIS alone.

In information retrieval it is typical to compare system performance experimentally. Therefore the comparative contribution type, and the system analytic contribution type, are common. Within the comparative type, LIS, computer science, and computer science jointly with social sciences scored notably higher compared to other disciplinary compositions, whereas within the system analytic type, LIS in collaboration with computer science scored clearly higher than LIS alone.

Thus, within information retrieval it is beneficial for LIS to join forces with computer science for greater research impact. This impact is likely associated with the use of mathematical and system analytic strategies in papers authored jointly by LIS and computer science, because they receive somewhat higher impact scores than some other disciplinary compositions, including LIS alone. This hints, that collaboration with computer scientists augments LIS by welcome methodical knowledge on information retrieval.

Within information seeking computer science in collaboration with social sciences received significantly higher impact scores compared to LIS. According to Järvelin and Vakkari (2022), survey and qualitative research strategies were the most popular strategies. In the use of survey, computer science in collaboration with social sciences produced significantly higher impact scores than LIS, LIS and social sciences and several other disciplinary compositions. This suggests that survey is an effective methodological choice for collaborating computer and social scientists in this subfield.

Within scientific communication, LIS with computer science received notably higher impact scores than several other disciplinary compositions. On the other hand, natural sciences and medicine received lower scores than several other compositions, including LIS. Järvelin and Vakkari (2022) have shown that in scientific communication, citation analytic, survey and case study strategies are the most common ones. Our findings indicate that within citation analytic strategy, LIS and computer science produce significantly higher impact scores than natural sciences and medicine. Thus, also scientific communication benefits from collaboration with computer science to produce results of high impact. This seems associated with the citation analytic strategy in joint articles by LIS and computer science. They receive higher impact scores than other corresponding articles by many disciplinary compositions.

In all, we wanted to find out, whether article content factors: the disciplinary composition, the study topic, research strategy, and the type of contribution are associated with the citation impact of the article. We found significant associations, discussed above, but less and weaker than we expected. This suggests that all strategies and contribution types are welcome if they are competently executed/produced. The type of strategy and contribution are not the master keys to explain the variation of citation impact. There is room for other content factors, such as novelty of research questions and design, quality of the data set and quality of methodology to explain the variance of citation impact. There is also room for other contextual factors, such as the standing of the authors, their institutions and of the journals within the academic community, to explain the variance. Some of this data is quite easy to collect, some very laborious – assessing novelty related factors being particularly notorious.

## 5.3 Limitations

The citation window for observing the number of citations received by papers citing the 2015 articles is limited. Therefore, the impact figures reflect early reactions to the articles in the research community. Moreover, the citation data is unqualified: "document $x$ cites article $y$" does not indicate what content categories in $y$ were cited by which in $x$, whether the citation is justified or important, or its stance. Consequently, the findings are less precise than is desirable, and their explanation power reduced. Third, we do not know, which discipline among the authorship is to be (dis)credited for a given citation. Fourth, while the dataset gives the disciplinary composition of the primary articles, there is no data on the discipline of the citing documents, thereby excluding some interesting research questions. Finally, while the primary articles in the dataset were published in volume 2015 issues of the journals, they may have been available online earlier or only after 2015. This may vary between journals and affect the citation window size in practice.

## 5.4 Further Studies

There are interesting possibilities for further studies based on, or by extending, the current data set. One is a qualitative analysis of innovativeness of study designs of a few lowly vs. highly cited articles of selected LIS topics. Another would classify the disciplines of the citing secondary and tertiary documents to allow the analysis of the flow of ideas from LIS studies. Of course, one could turn the analytic eye toward references and the influx of ideas to LIS. Finally, the corresponding datasets for the years 1995 and 2005 could be extended with reference/citation data to foster longitudinal analysis of the flow of ideas. One may also examine, whether the contextual factors: author names, journal titles, institution names, and country names, etc., have stronger relationship with citation impact than the content factors analyzed herein. Regarding methodological development, the available datasets could be used as ground truth for the development of data mining / machine learning methods for automatic classification of content dimensions and contribution recognition (Goh *et al.,* 2020).

# 6 CONCLUSION

We have analysed the citation impact of scholarly articles published in 31 leading international LIS journals in 2015. We focused on the degree to which authors' disciplinary composition together with content characteristics of articles affects their citation impact. An article's citation impact was based on the number and authority of citations received but normalised for outliers and LIS subfield sizes present in the dataset. The citations received by the articles were traced from 2015 to May 2021. Citing document authority was measured by the citations they received within the same citing window.

Our overall finding was that authors' disciplinary composition is significantly associated with citation impact scores. The differences in the scores between disciplinary compositions appeared typically within information retrieval and scientific communication. In both topics LIS and computer science jointly received significantly higher citation scores than many disciplines like LIS alone or humanities in information retrieval; or natural sciences and medicine, or social sciences alone in scientific communication.

The main limitations of the present study – the unqualified citation data; which discipline brought which citation; which disciplines do the citing documents represent – suggest ideas for further studies to raise the level of explanation of citation impact. On the other hand, this study is original in allowing joint analysis of content, authorship composition, and impact.

## Appendix I

| Journal Name 2015 | Vols | No of Arts |
|---|---|---|
| ACM Transactions on Information Systems | 33(1) -34(1) | 27 |
| Aslib Journal of Information Management (formerly: Aslib Proc.) | 67 | 36 |
| College and Research Libraries | 76 | 57 |
| Information & Culture  (formerly: Libraries & Culture ) | 50 | 24 |
| Information Processing and Management | 51 | 65 |
| Information Research | 20 | 46 |
| Information Retrieval | 18 | 21 |
| Information Services & Use | 35 | 27 |
| Information Technology and Libraries | 34 | 19 |
| International Information & Library Review | 47 | 10 |
| International Journal of Information Management | 35 | 71 |
| Journal of Documentation | 71 | 71 |
| Journal of Education for Library and Information Science | 56 | 27 |
| Journal of Information Science | 41 | 57 |
| Journal of Librarianship and Information Science | 47 | 32 |
| Journal of Library Administration | 55 | 22 |
| Journal of the Association for Information Science & Tech | 66 | 196 |
| Library & Information History  (formerly: Library History) | 31 | 11 |
| Library and Information Science Research | 37 | 40 |
| Library Collections, Acquisitions, and Technical Services | 39 | 11 |
| Library Quarterly | 85 | 24 |
| Library Resources and Technical Services | 59 | 15 |
| Library Trends | 63 | 47 |
| Libri | 65 | 24 |
| New Review of Information Networking | 20 | 27 |
| Online Information Review | 39 | 52 |
| Program | 49 | 24 |
| Reference & User Services Quarterly (formerly: Reference Quart.) | 54(3)-55(2) | 12 |
| Scientometrics | 102-105 | 345 |
| The Electronic Library | 33 | 70 |
| The Indexer | 33 | 30 |
| **TOTAL** | | **1540** |

# Appendix II – Classifications for Content Analysis

## LIS Topics By Main Topic

### *Research on L&I context and services*

010  the professions
020  library history, history of L&I institutions
030  publishing

100  education in LIS Studies
200  methodology
300  analysis of LIS discipline

410  document delivery
420  collections
430  Information or reference service
440  user education or information literacy education
450  L&I service buildings
460  administration or planning
470  automation or digital libraries
480  other L&I services
490  several interconnected activities

800  other aspects of LIS

### *Research in information storage and retrieval*
510  metadata / cataloguing
520  classification and indexing
531  text retrieval
532  retrieval methods in other media
533  web retrieval methods
534  social media retrieval
540  digital information resources
550  interactive (user-oriented) IR
560  other aspects of IR)

### *Research on information seeking*
610  information dissemination
620  use or users of channels or sources of information
630  use of L&I services
641  task-based information seeking
642  other type of information seeking
650  information use
660  information management

### *Research on scientific comm*
710  scientific / professional publishing
720  citation patterns and structures
730  web-metrics
740  other aspects of sci/prof communication

## RESEARCH STRATEGY

### *Empirical*
11    historical
12    survey
13    qualitative
14    evaluation
15    case or action research

16    content or protocol analysis
17    citation analysis
18    other bibliometric
21    secondary analysis
22    experiment
29    other empirical strategy

*Conceptual*
31    verbal argumentation
32    concept analysis

*Other non-empirical*
40    mathematical or logical
50    system analysis and design
60    literature review
80    bibliographic
90    other strategy
00    not applicable

**TYPE OF INVESTIGATION**

*Empirical*
11    descriptive
12    comparative
13    explanatory

*Non-empirical*
20    conceptual
30    theoretical
40    methodological
50    system design

*Other  contributions*
90    other type
00    not applicable

# Appendix III – Affiliation-based Discipline Classes

**MainClass & Sample Subclasses (Chang 2018)**

*Business and Economics*
    Business
    Economics
    Management

*Computer sciences*
    Computer science and engineering
    Information systems and HCI

*Engineering*
    Engineering
    Architecture
    Energy

*Humanities*
    Humanities
    Literature
    Arts
    Anthropology

Linguistics
Philosophy and religion
History

*Library and information Science (LIS)*
Documentation
information Science
Library Science

*Medicine*
Medicine
Nursing
Health science

*Natural Sciences*
General science
Physics
Mathematics
Biology
Agriculture
Chemistry
Zoology
Botany

*Social sciences*
Education
General social science
Communication
Law
Psychology
Sociology
Political science
Tourism

*Other*
*any other non-fitting or unknown discipline*